\documentclass[conference]{IEEEtran}
\IEEEoverridecommandlockouts
\usepackage{cite}
\usepackage{amsmath,amssymb,amsfonts}
\usepackage{algorithmic}
\usepackage{graphicx}
\usepackage{textcomp}
\usepackage{xcolor}
\usepackage{subcaption}
\usepackage{authblk}

\usepackage{array}

\def\BibTeX{{\rm B\kern-.05em{\sc i\kern-.025em b}\kern-.08em
    T\kern-.1667em\lower.7ex\hbox{E}\kern-.125emX}}
    
\begin{document}

\title{Robust Ambient Backscatter Communications with
Polarization Reconfigurable Tags}
\author[1,2]{Romain~Fara} 
\author[2]{Dinh-Thuy~Phan-Huy} 
\author[3]{Abdelwaheb~Ourir}
\author[1]{Marco~Di Renzo} 
\author[3]{Julien~de Rosny}
\affil[1]{Laboratoire des Signaux et Systemes, University Paris Saclay, CNRS, CentraleSupelec, Gif Sur Yvette, France}
\affil[2]{Orange Labs Networks, Chatillon, France} 
\affil[3]{ESPCI Paris, PSL University, CNRS, Institut Langevin, Paris, France \authorcr{romain.fara@orange.com}}
\maketitle   
\begin{abstract}

Ambient backscatter communication system is an emerging and promising low-energy technology for Internet of Things. In such system, a device named tag, sends a binary message to a reader by backscattering a radio frequency signal generated by an ambient source. Such tag can operate without battery and without generating additional wave. However, the tag-to-reader link suffers from the source-to-reader direct interference. In this paper, for the first time, we propose to exploit a “polarization reconfigurable” antenna to improve robustness of the tag-to-reader link against the source-to-reader direct interference. Our proposed new tag sends its message by backscattering as an usual tag. However, it repeats its message several times, with a different radiation pattern and polarization, each time. We expect one polarization pattern to be better detected by the reader.  We show by simulations and experiments, in line-of-sight and in richly scattering environment, that a polarization reconfigurable tag limited to 4 polarization directions outperforms a non-reconfigurable tag and nearly equals an ideally reconfigurable tag in performance.
\end{abstract}


\section{Introduction}
Recent development of Internet of Things (IoT) has massively increased the number of connected devices. At the same time, despite the energy efficiency improvement brought by every mobile network generation, the energy consumption keeps increasing due to the fast growth of the number of devices \cite{gati_key_2019}.

Recently, ambient backscatter (AmB) principle \cite{liu_ambient_2013} has been proposed for low-energy consumption communication. In AmB system, a Radio-Frequency (RF) tag transmits a binary message to a RF reader without battery and without generating additional wave. The tag must be illuminated by a RF ambient source (such as a TV tower, Wi-Fi hot-spot or 5G base station ...). Basically, the tag switches between two states: a backscattering state, in which it backscatters the ambient signal, and a transparent state, in which it has a weaker effect on the ambient signal. The two distinct states code for bit “1” and “0”, respectively. The simplest implementation of a tag is a dipole antenna that switches between two different load impedances: a null impedance (the two branches of the dipole are short-circuited) or an infinite impedance (the two branches of the dipole are open-circuited). Such tag does not generate any additional RF wave and can therefore operate without battery. An energy-harvesting device is then sufficient to power a RF switch and a low power micro-controller. On the RF reader side, the simplest receiver is an energy-detector that compares the current received level of power to a threshold (the time-windowed received power for instance) to determine in which state the tag is and to deduce the sent bits. The performance of the energy-detector increases with the SNR contrast, i.e. the difference between the values of the receive signal-to-noise ratio (SNR) in the first state and in the second state, respectively.

Due to its low energy consumption, the AmB principle has been identified as a promising technology for IoT \cite{zhang_green_2019}. However, the tag-to-reader link suffers from a source-to-reader direct interference in many different ways. First, in average, the tag-to-reader signal is weak compared to the source-to-reader direct interference. This is due to the fact that a fraction of the incident ambient signal is backscattered and spread in many directions. Secondly, a tag in a deep fade of the ambient signal is invisible to the reader. This may happen, typically, in a richly scattering environment. Thirdly, even in line-of-sight (LOS), and even when the tag-to-reader signal is strong, the SNR contrast can be close to zero for some locations of the tag. Indeed, there are locations where the tag-to-reader signal and the source-to-reader direct interference combine in such a way that the received signals in the two distinct states are distinct in phase but equal in amplitude. In these locations, the SNR contrast is null and the energy-detector performance is poor. \cite{rachedi_demo_2019} shows by simulation and experiments, that SNR contrast deep fades occur when the tag is located on ellipses that have the source and the reader as foci. These ellipses are regularly spaced by half a wavelength. Such SNR contrast deep fades may also occur in scattering environments. Finally, even if the tag is out of such locations, if the polarizations of the tag, reader and source do not match, the performance can also be poor \cite{van_huynh_ambient_2018}.

Channel polarization has been studied extensively for RFID tags and RFID readers in \cite{mhanna_statistical_2013}. To guarantee a minimum polarization match, a circular polarized antenna at the reader side has been proposed in \cite{hebib_antennes_2011}. However, ambient sources may not use such type of antenna systematically.

More recently, a tag based on a compact reconfigurable antenna has been proposed to reach higher data rates in AmB systems in \cite{kokar_first_nodate}. The antenna switches between 4 radiation patterns with distinct dominant linear polarizations. This “polarization reconfigurable” tag attains log(2)=2 bits/switching period instead of 1 bit/switching period (that would be obtained with a 2-state tag). 

\begin{figure}
    \centering
    \includegraphics[trim=0cm 0cm 0cm 0cm,clip,width=\columnwidth]{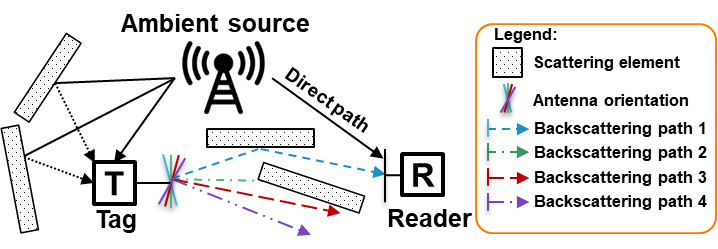}
    \caption{PR tag illuminated by ambient signal in a richly scattering environment, and backscattering its message to the reader, several times, with different polarizations, to improve robustness.}
    \label{fig:principle}
\end{figure}

In this paper, for the first time, we propose to exploit such “polarization reconfigurable” tag to improve robustness of the tag-to-reader link against the source-to-reader direct interference. As illustrated in Fig. \ref{fig:principle}, we consider a polarization reconfigurable (PR) tag able to switch between several polarizations. However, in this paper, our tag does not communicate by switching between polarization. More precisely, as any standard two-state tag, the PR tag sends its message by switching its antenna between the backscattering state and the transparent state (i.e. by switching the antenna between two different load impedances). However, the PR tag sends the same message several times, with a different configured radiation pattern and the corresponding different polarization, each time. We expect that among all the used polarizations, one will improve the energy-detector performance. 

To make an initial assessment of the benefit of such PR tag, we propose a very simple model of the PR tag that can easily be studied theoretically, numerically using a simulation software (4NEC2) of computation electromagnetics based on the Method of Moments, and experimentally using dipoles. We model a PR tag that can switch between $Npol$ radiation patterns (and corresponding polarizations), by a mechanically rotating dipole that switches between $Npol$ different orientations. The dipole is connected to a load impedance with two different values to create the backscattering and the transparent state. Three types of PR tags are considered in our studies: \begin{enumerate}
    \item an “ideal” polarization reconfigurable (IPR) tag able to switch between any directions of polarization,
    \item a more realistic 4-polarization reconfigurable (4PR) tag limited to 4 directions of polarization,
    \item a non-reconfigurable (NR) tag with a fixed direction of polarization.
\end{enumerate}
{}
The paper is organized as follow: Section \ref{system_model} presents our system model. Section \ref{LOS} presents a theoretical and simulation based performance analysis in LOS and derives a simple analytical method to determine the best direction of polarization. Section \ref{richly_scattering_environment} presents our simulation and experimental results for a richly scattering environment and Section \ref{conclusion} concludes this paper.

\section{System Model}
\label{system_model}
\subsection{Propagation and Environment Model}

\begin{figure}[h]
    \centering
    \includegraphics[trim=0cm 0cm 0cm 0cm,clip,width=\columnwidth]{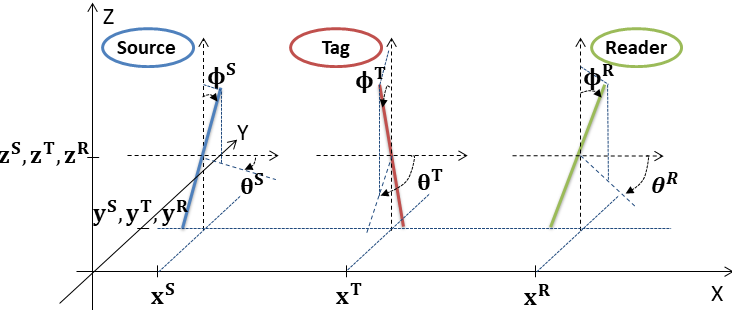}
   \caption{3D Model of the system}
   \label{3dmodel}
   
\end{figure}

In this paper, we consider a system composed of a source, a PR tag and a reader. The source radiates an ambient signal at the carrier frequency $f$ with the corresponding wavelength $\lambda$.
Source and reader are equipped with a half-wavelength linearly polarized dipole antenna of length $l^D=\lambda/2$. In this model, we neglect the radius of all the conductors (thin-wire approximation).

The tag’s antenna switches between $Npol $ radiation patterns with distinct dominant linear polarizations. Note that $Npol=1$, $Npol=4$, $Npol=\infty$ corresponds to NR tag, 4PR tag and IPR tag, respectively. The tag is modelled as a mechanically rotating half-wavelength dipole antenna with linear polarization. Thanks to this extremely simple model, we can focus our study on the impact of linear polarization and evaluate the best gains that can be drawn from polarization reconfiguration. Future studies will take into account the joint impacts of the antenna gain, directivity and polarization, together.

Sources of propagation channel diversity, such as scatterers and reflectors, are taken into account in this model as follows. A scatterer is modeled as a conductive line defined by its length $l^{SC}$. We define $N^{SC}$ as the number of scatterers. $N^{RP}$ infinite reflective planes, such as a ground plane for instance are used to model reflectors.

The positions and orientations of the dipole antennas of source, tag and reader  are defined as follows: ($x^S$, $y^S$, $z^S$), ($x^T$, $y^T$, $z^T$) and ($x^R$, $y^R$, $z^R$) are the Cartesian coordinates of the centers of the dipole antennas of source, tag and reader, respectively; ($\phi^S$, $\theta^S$), ($\phi^T$, $\theta^T$) and ($\phi^R$, $\theta^R$), are the orientation angles of the dipole antennas of source, tag and reader, respectively (see Fig. \ref{3dmodel}). Each scatterer is randomly positioned to maintain a distance to each dipole, $D^{SC-X}>\lambda$ and a distance to the reader $D^{SC-R}<10\lambda$. Also the distances between any pair of dipoles are chosen higher than half of the wavelength. Hence, in this paper, scattering, reflection and backscattering effects are considered. Near-field coupling effects are out of the scope of this study.

The load impedances of the dipole antennas of the reader and the source are defined hereafter. 
The dipole antenna of the reader is connected to a receiver with an equivalent load impedance $Z^R$ that is fixed and adapted to the antenna. 
The tag switches between two load impedances to communicate and rotates mechanically its dipole antenna (according to $Npol$ available orientations) to emulate polarization reconfiguration. More precisely, the tag sends its message several times, with a different orientation, each time. For a given orientation of the dipole, the tag sends its message by switching between two states: the backscattering state and the transparent state. Each state is obtained by connecting the dipole antenna to a distinct load impedance. The backscattering state is obtained by short-circuiting (ON) the dipole antenna strands, i.e. by setting the tag load impedance to $0$. The transparent state is obtained by open-circuiting (OFF) the dipole antenna strands, i.e. by setting the tag load impedance to $\infty$. These two states have different impacts on the propagation between the source and the reader.

\subsection{Performance metrics}

At the reader side, the detector measures the voltage on the dipole antenna port, induced by the total signal received by the antenna. We denote $V^{ON}, V^{OFF}$ the voltage measured at the dipole antenna port of the reader, when the tag is backscattering and when the tag is transparent, respectively. From the voltage value, the detector deduces for each state the corresponding power $P^{ON}$ and $P^{OFF}$.
We define $\Delta P$ as the difference of received power between the two states of the tag: 
\vspace{-0.2cm}
\begin{equation}
    \Delta P=|P^{ON}-P^{OFF}|.
\end{equation}
We denote $P^{noise}$ as the receiver noise power level. Using these notations, we define the SNR contrast ($\Delta SNR$) as:
\begin{equation}
    \Delta SNR =\frac{\Delta P}{P^{noise}}.
\end{equation}
We define $SNR^{Tx}$ as the power transmitted by the source divided by the received noise power of the reader. To allow a fair comparison between all types of PR tags, transmitted power and received noise power are considered as fixed for the study. 
We also define $SNR^{captured}$ that reflects the average $SNR$ of the signal captured by the reader when the tag is transparent, $SNR^{captured}=\mathbb{E}_{x^T,y^T}[\frac{P^{OFF}}{P^{noise}}]$.

As the reader uses an energy detector to decode the signal, the bit error rate ($BER$) is given by \cite{rachedi_demo_2019} : 
\begin{equation}
    BER=\frac{1}{2} erfc(\Delta SNR).
    \label{eq:ber}
\end{equation}{}
To achieve a given quality of service (QoS) associated to a $BER^{target}$, $\Delta SNR$ needs to be higher than the value $\Delta SNR^{target}$.
Also by comparing the measured $\Delta SNR$ to the $\Delta SNR^{target}$, we derive the probability of outage of the proposed system, this probability being assessed over a target coverage area.

\subsection{4NEC2 Simulator}

In this paper, the simulations have been made using 4NEC2 software simulator. 4NEC2 is a numerical electromagnetic code (NEC) based on the Method of Moments. The tool allows us to fully model and configure the source, the tag, the reader, $N^{SC}$ scatterers, a ground plane, the load impedances and the physical characteristics of the dipole antennas and the conductive lines. This tool allows us to 1) configure the state of the tag (transparent or backscattering), by changing the load impedance; 2) measure the resulting voltage at reader port, induced by the electric field received by the reader.

\section{LOS environment analysis}
\label{LOS}

In this section, we consider LOS propagation, i.e. with $N^{SC}$ and $N^{RP}$ equal to $0$. Only the source, the tag and the reader are considered. The carrier frequency $f$ is set to 2.4 GHz.
In this section, we limit the study to an IPR tag that allows us to find the best polarization for the tag and some upper bound performance. 4PR and NR tags derive from IPR tag as they involve a limited number of polarization directions. Comparison of the performance of these tags is presented in the next section.
\subsection{Optimum polarization selection based on a simple analytical model (OPSSA) }
\label{LOS_theory} 
We propose an optimum polarization selection based on a simple analytical model (OPSSA) scheme. We use simplifying assumptions, to find the orientation of the linear polarization that maximizes $\Delta SNR$. We assume that the source, the tag and the reader have a perfectly linear polarization. Under these assumptions, the direction of the electric field is given by the orientation of the dipole antennas. Let $\Vec{\mathbf{S}}$ be the normalized electric field vector and let $\Vec{\mathbf{T}}$ and $\Vec{\mathbf{R}}$ be the unitary vectors giving the orientations of the dipoles of the tag and the reader, respectively.
For a given position and orientations of the dipole antennas, we approximate the direct source-to-reader signal, $S^{direct}$ (for a normalized electric field $\Vec{\mathbf{S}}$), as the result of the projection of $\Vec{\mathbf{S}}$ over $\Vec{\mathbf{R}}$. The source-to-tag-to-reader signal ($S^{back}$) is approximated as the projection of $\Vec{\mathbf{S}}$ over $\Vec{\mathbf{T}}$, then on $\Vec{\mathbf{R}}$:
\begin{equation}
    S^{direct}=\Vec{\mathbf{S}}\cdot \Vec{\mathbf{R}},
\end{equation}
\begin{equation}
    S^{back}=(\Vec{\mathbf{S}}\cdot \Vec{\mathbf{T}})(\Vec{\mathbf{T}}\cdot \Vec{\mathbf{R}}),
\end{equation}
where $\cdot$ is the dot product. With these notations and for a fixed vertical source ($\phi^S=0$) we obtain the following optimal solution:
\begin{equation}
    \begin{cases}
        \theta^T=\theta^R \ [2\pi],\\
        \phi^T=\frac{\phi^R}{2} \ [\frac{\pi}{2}].
    \end{cases}
    \label{eq:theoretical_model}
\end{equation}
Results from Equation \ref{eq:theoretical_model} show that there exists a best orientation for the tag. Therefore, an IPR tag will outperform a NR tag.
The derived optimum orientation of the tag can be interpreted as follows: the best orientation of the IPR tag is obtained when the tag simultaneously maximizes the received signal from the source and maximizes the backscattered signal to the reader. This is obtained when the angle between the source and the tag equals the angle between the tag and reader.
\subsection{ 4NEC2 simulation based validation}
We validate this OPSSA approach by simulation. 
We recall that the 4NEC2 tool takes into account the following elements that are neglected in the simple analytical model: the true wire length, coupling between elements, propagation LOS and radiating diagram.
For a given set of source, tag and reader locations and for a given set of reader orientations, we determine numerically by simulation, through an exhaustive search, the orientation of the tag that maximizes $\Delta SNR$. The numerical search is performed over a reduced number of angles due to the symmetry: ($\phi^R$,$\theta^R$) in$([0,10, \dots, 90],[0,10, \dots,90])$ and ($\phi^T$,$\theta^T$) in$([0,1, \dots, 90],[0,1, \dots,180])$ in degrees. For a given reader orientation, once we have found the optimum tag orientation, we compare two unitary vectors giving the best orientation of the tag according to two different methods: 
\begin {itemize}
\item ($\Vec{\mathbf{T}}^{best-th}$) that is obtained through the OPSSA method
\item ($\Vec{\mathbf{T}}^{best-simu}$) that is obtained through exhaustive search.
\end {itemize}

\begin{figure}[h]
   \includegraphics[trim=0cm 0cm 0cm 0.8cm,clip,width=\columnwidth]{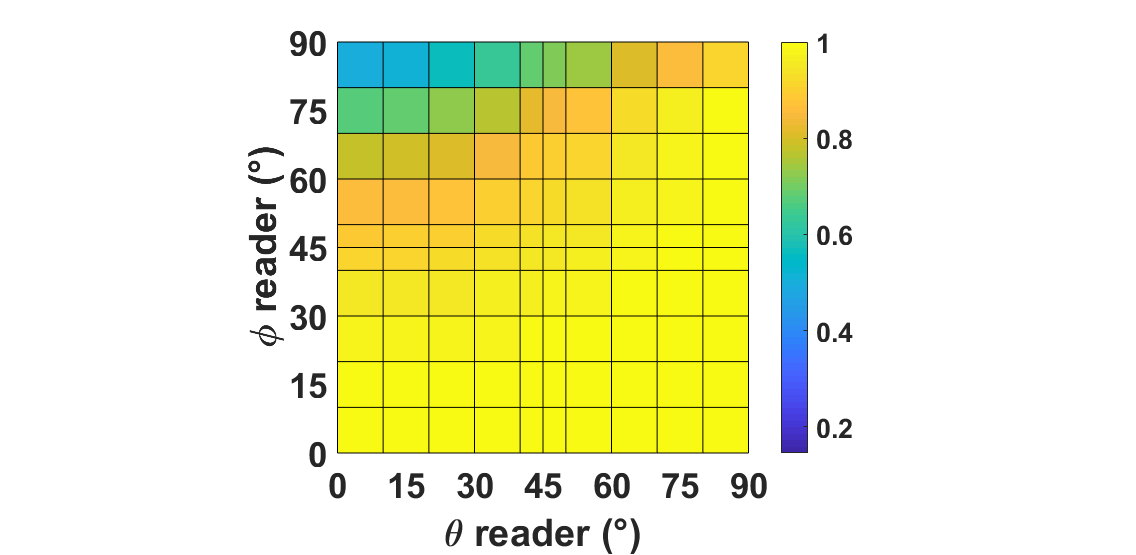}
   \caption{Comparison of the best orientation obtained with OPSSA scheme and exhaustive search (dot product of $\Vec{\mathbf{T}}^{best-th}$ and $\Vec{\mathbf{T}}^{best-simu}$).}
   \label{fig:theory_vs_simu}
\end{figure}

Fig. \ref{fig:theory_vs_simu} illustrates the dot product between $\Vec{\mathbf{T}}^{best-th}$ and $\Vec{\mathbf{T}}^{best-simu}$. A high dot product (close to 1) means that the polarization found with the OPSSA scheme matches the one obtained by exhaustive search. Fig. \ref{fig:theory_vs_simu} shows polarizations match of more than 80\% for reader orientation $\theta^R >50^\circ$ or $\phi^R<45^\circ$. The OPSSA approach is hence, valid in most cases. The cases where the model is not valid correspond to the source being out of the “donut” diagram of the reader dipole and are therefore of low interest.

\begin{figure}[h]
    \centering
   \includegraphics[trim=0cm 0cm 0cm 0cm,clip,width=0.8\columnwidth]{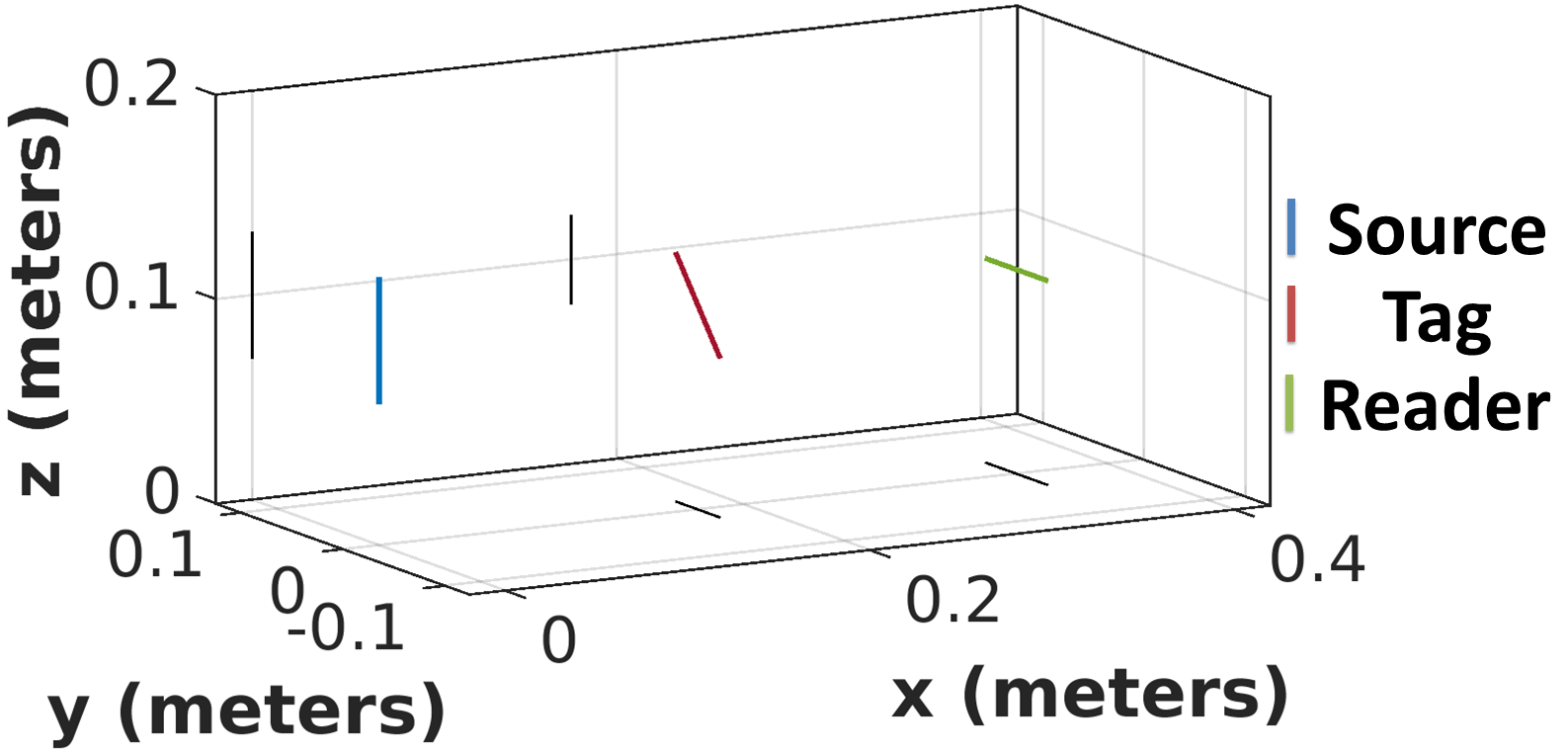}
\caption{Example of optimum angle for a given source $(\theta^S,\phi^S)=(0^\circ,0^\circ)$ and reader $(\theta^R,\phi^R)=(90^\circ,90^\circ)$ configuration.}
    \label{fig:s0_t45_r90_config}
\end{figure}

To conclude this section, in LOS, the IPR tag has to be “half-way” between the reader and the source, in terms of orientation. In other terms, the orientation of the IPR tag has to be the average between the orientation of the source and the orientation of the reader.

Let us observe the advantage of a PR tag in the worst case scenario, where, the source and the reader are orthogonal to each other, i.e. the source is 90° from the reader. In this case, the reader cannot receive any signal from the source. The current study shows that the IPR tag is better detected by the reader, when it is transmitting its message, with the direction of its linear polarization being at 45° from the source and 45° from the reader, as shown in Fig. \ref{fig:s0_t45_r90_config}.

In a real environment, in addition to the LOS propagation paths studied in this section, additional propagation paths exist (due to scatterers and reflectors) that will be studied in the next section.

\section{Comparison of the tags in richly scattering environment}
\label{richly_scattering_environment}
In this section, we consider a richly scattering environment, with a large number of scatterers and few reflective planes. The carrier frequency $f$ is set to 2.4 GHz. We consider the three different types of tags, we recall that 4PR and NR tags are modelled by setting $Npol=4$ and $Npol=1$, respectively. We simulate the IPR tag with $Npol=81$.

The simulation models $N^{SC}=20$ scatterers and one ground plane ($N^{RP}=1$). The waves travel from the source to the reader through the LOS path and multiple additional non line-of-sight (NLOS) paths due to scatterers in proximity of the reader and the ground plane. Whereas the linear polarization of the wave remains unchanged along the LOS path, it may change on the NLOS paths. Hence, at the reader side, the source-to-reader signal results from the combination of incident waves with distinct linear polarizations. We expect the linear polarization of the LOS path to be dominant in this combination, as the LOS path is expected to be the stronger in power.
\begin{table}[ht]
\caption{System model parameters}
\begin{center}
\begin{tabular}{|c|c|c|c|}
\hline
\textbf{\textit{Parameters}} & \textbf{\textit{Details}}& \textbf{\textit{Value}}& \textbf{\textit{Subhead}} \\
\hline
$(x^S,y^S,z^S)$& Source localization & $(0,0,0.3)$& m \\
\hline
$(\phi^S,\theta^S)$& Source Orientation& $(0,0)$& deg \\
\hline
$(x^R,y^R,z^R)$& Reader localization & $(100,0,0.3)$& m \\
\hline
$(\phi^R,\theta^R)$& Reader Orientation& $(90,90)$& deg \\
\hline
$(x^T,y^T,z^T)$& Tag localization & $(x^T,y^T,0.3)$& m \\
\hline
$l^{SC}$& Length of scatterers& $\lambda/2$&m \\
\hline

$Z^{R}$& Reader load impedance& $50$&$\Omega$ \\
\hline

$BER^{target }$& Target Bit error rate & $10^{-2}$& \\
\hline
$\Delta SNR^{target}$& Target contrast value & $3.4$& dB \\
\hline

\end{tabular}
\label{table:param_table}
\end{center}
\end{table}

\subsection{Visualization results}

In this section, we propose to visualize the best linear polarization of the tag obtained using 4NEC2 simulations. Source and reader orientations are fixed and orthogonal to each other, as it is a worst case scenario in terms of received SNR. 
We analyze the performance of the communication as a function of tag’s coordinates $(x^T,y^T)$ , i.e. we draw spatial 2D map of $\Delta SNR$. Other parameters of the simulation with fixed values, are detailed in Table \ref{table:param_table}. 
\begin{table}[ht]
\caption{LOS and scattering environment configuration}
\begin{center}
\begin{tabular}{|c|c|c|}
\hline
 &LOS &Scattering \\
\hline
NR tag in the worst orientation & LOS-NR-Worst & SCAT-NR-Worst \\
\hline
NR tag in the best orientation & LOS-NR-Best & SCAT-NR-Best \\
\hline
4PR tag & LOS-4PR & SCAT-4PR \\
\hline
IPR tag& LOS-IPR & SCAT-IPR \\
\hline
\end{tabular}
\label{table:config_table}
\end{center}
\end{table}
We study the three types of tag for different configurations of orientation and environment (LOS or with scattering). Fig. \ref{fig:maps} illustrates $\Delta SNR$ maps of some of the listed configurations in Table \ref{table:config_table} for a given $SNR^{Tx}=110dB$. The spatial maps of $\Delta SNR$ show the locations where the tag can be detected by the reader with the target QoS ($\Delta SNR>\Delta SNR^{target}$). Light colors (yellow or red) indicate where QoS can be achieved and dark colors (blue) indicate locations where QoS cannot be reached. Each subfigure of Fig. \ref{fig:maps} is accompanied by a “\textit{velvet carpet}” illustrating the tag’s best orientation depending on the tag’s location. Each position on the velvet carpet corresponds to a tested position of the tag, in space. The thread of the carpet at a given position, illustrates the best orientation of the tag, for this considered position. The velvet carpet illustrations on Fig. \ref{fig:maps}-a,c,d show that the orientation is uniform for NR tags and Fig. \ref{fig:maps}-b,e,f show that the orientation is non-uniform for PR tags. 

First, we show in Fig. \ref{fig:maps}-a and in Fig. \ref{fig:maps}-b the results obtained for the configuration in LOS environment from Section \ref{LOS}. Fig. \ref{fig:maps}-a illustrates the $\Delta SNR$ for the optimum orientation in LOS: $\theta^T=90^{\circ}$ and $\phi^T=45^{\circ} $ from \ref{LOS_theory}. Fig. \ref{fig:maps}-b is the contrast map, in LOS, for the IPR tag, detailed hereafter. We observe that a PR tag has limited effects in LOS configuration and with the source and the reader in cross-polarization. We then consider the configuration in the scattering environment.

\subsubsection{NR tag}: To provide a reference for comparison with PR tags, we study the performance of the NR tag when its fixed orientation is, “by chance”, the best in average. We determine numerically this best orientation. We observe that even in a scattering environment (Fig. \ref{fig:maps}-c), we obtain the same best orientation as  in a LOS environment. Indeed, the optimum orientation in LOS, obtained by applying Equation \ref{eq:theoretical_model}, to parameters listed in table \ref{table:param_table} is $\theta^T=90^{\circ}$ and $\phi^T=45^{\circ} $. This is due to the fact that our studied scattering environment is close to a LOS environment. For comparison, Fig. \ref{fig:maps}-d illustrates the performance of the NR tag when its fixed orientation is, “out of luck”, the worst in average ($\theta^T=90^{\circ} $ and $\phi^T=90^{\circ}$), i.e., for which a very small amount of backscattered signal can be detected by the reader.

\subsubsection{4PR tag}
\label{limited_reconfigurable_tag}
The compact reconfigurable antennas presented in \cite{kokar_first_nodate} have 4 patterns with distinct dominant linear polarization directions. The 4PR tag is a simplified model of such existing antennas. The angles of the 4 polarizations are set to $(\phi^T,\theta^T) =\{(0,90),(45,90),(90,90),(135,90)\}$ and correspond to the main polarization directions of one of the antennas from \cite{kokar_first_nodate}. SNR contrast maps are computed for the 4 polarizations of the tag. 

\subsubsection{IPR tag}
We simulate the IPR tag with $Npol=81$. The orientations of the IPR tag ($\phi^T$,$\theta^T$) are uniformly distributed in $([0,180],[0,180])$ in degrees. We compute SNR contrast maps for each of $Npol$ polarizations. 

\begin{figure}[h]    
\centering
    \includegraphics[trim=0cm 0cm 0cm 0cm,clip,width=\columnwidth]{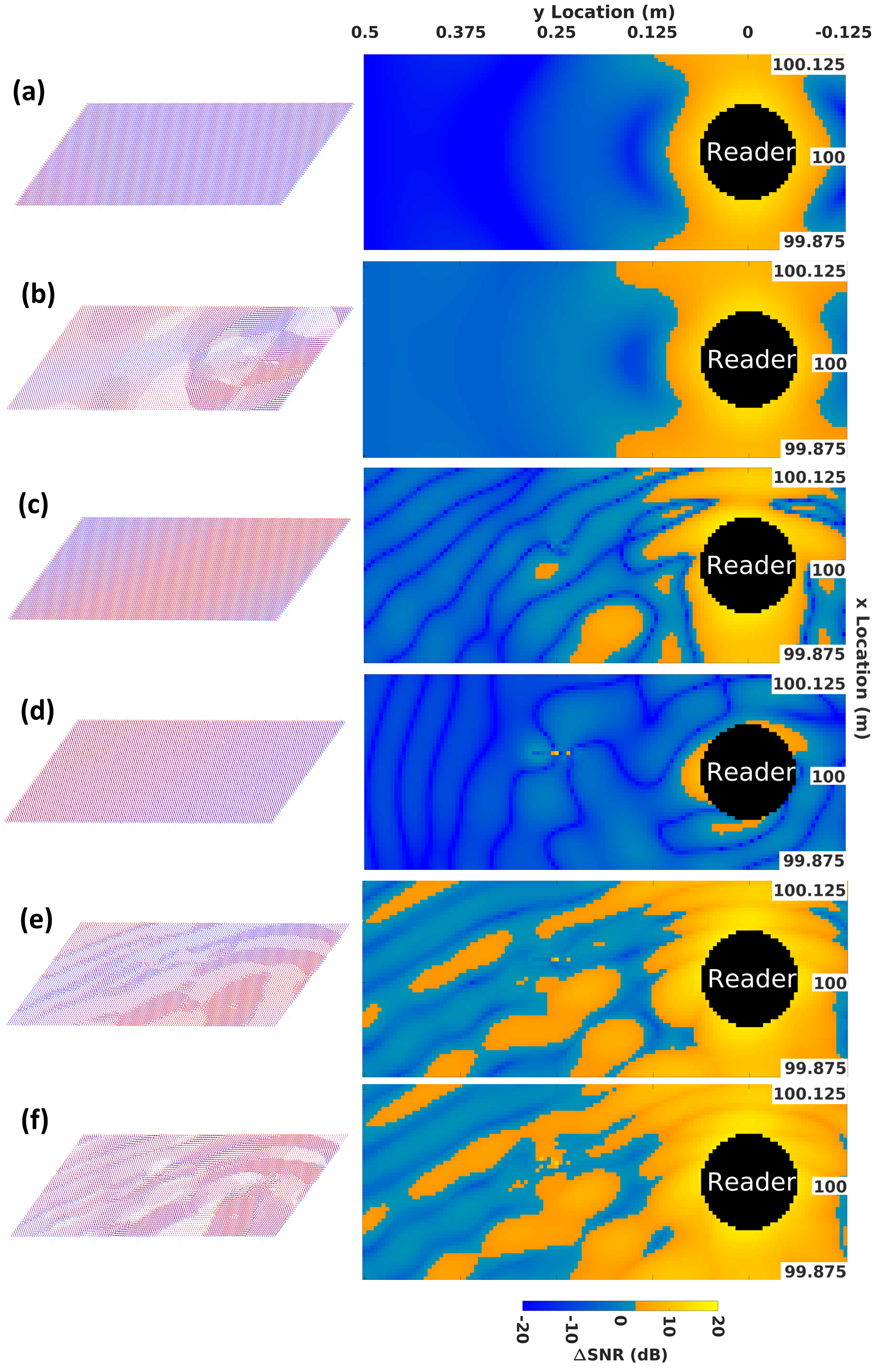}
   \caption{ $\Delta SNR$ maps showing the tag location that guarantees QoS (in orange) for the configurations : (a) LOS-NR-Best, (b) LOS-IPR, (c) SCAT-NR-Best, (d) SCAT-NR-Worst, (e) SCAT-4PR, (f) SCAT-IPR. In bottom right corner of each subfigure, the tag orientation map is illustrated.}
    \label{fig:maps}
\end{figure}

For the IPR and 4PR tags, we determine the polarization of the tag that maximizes the SNR contrast for every location of the tag based on the $Npol$ computed maps of each tag. We obtain the corresponding optimum SNR contrast maps illustrated in the Fig. \ref{fig:maps}-e,f

As expected, Fig. \ref{fig:maps}, shows that the IPR tag outperforms all types of tags. We also observe that the more realistic 4PR tag, with 20 times less available polarizations than the IPR tag, is close to the IPR tag, in performance. The 4PR tag outperforms the NR tag, even if the NR tag is ‘by chance’ using the optimum orientation of the LOS system as a fixed orientation. Compared to the NR tag, the 4PR tag is more robust to the impact of scatterers and polarization mismatch. We have visualized the impact of each PR tag on $\Delta SNR$ maps for a given $SNR^{Tx}$, then we study the performance of the tag as a function of $SNR^{Tx}$.

\subsection{Outage probability analysis}

The previous section has shown that a PR tag can improve the performance of the AmB system in terms of SNR contrast. 
In this section, we numerically evaluate the outage probability for each configuration of the Table \ref{table:config_table}. The probability is computed over a target coverage area, i.e. over tag’s locations (in meters) defined by: $x^T=x^R+\alpha\times0.001$ and $y^T=y^R+\beta\times0.001$, with $0.5\lambda<D^{T-R}<3\lambda$ and $\alpha,\beta \in \mathbb{Z}$, where $D^{T-R}$ corresponds to the euclidian distance between, the tag and the reader. 
$\Delta SNR$ is calculated as a function of $SNR^{Tx}$, for every tag location and orientation and for a given environment configuration of the source, the reader, the scatterers and the reflective planes. We compute the outage probability, for the three types of tags.
\begin{figure}[h] 
    \centering
   \includegraphics[trim=0cm 0cm 0cm 0cm,clip,width=\columnwidth]{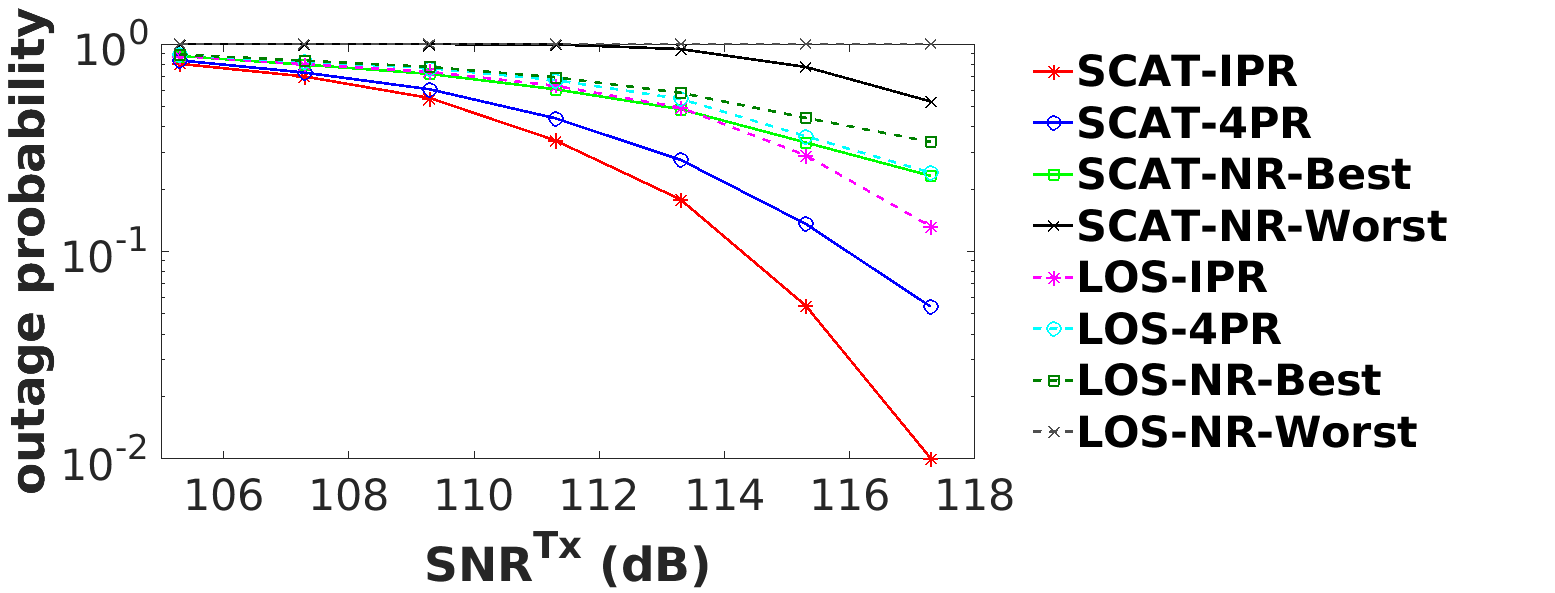}
   \caption{Outage probability simulated as a function of $SNR^{Tx}$.}
   \label{simulation_results}
\end{figure}
\begin{figure}[h]
    \centering
   \includegraphics[trim=0cm 0cm 0cm 0cm,clip,width=\columnwidth]{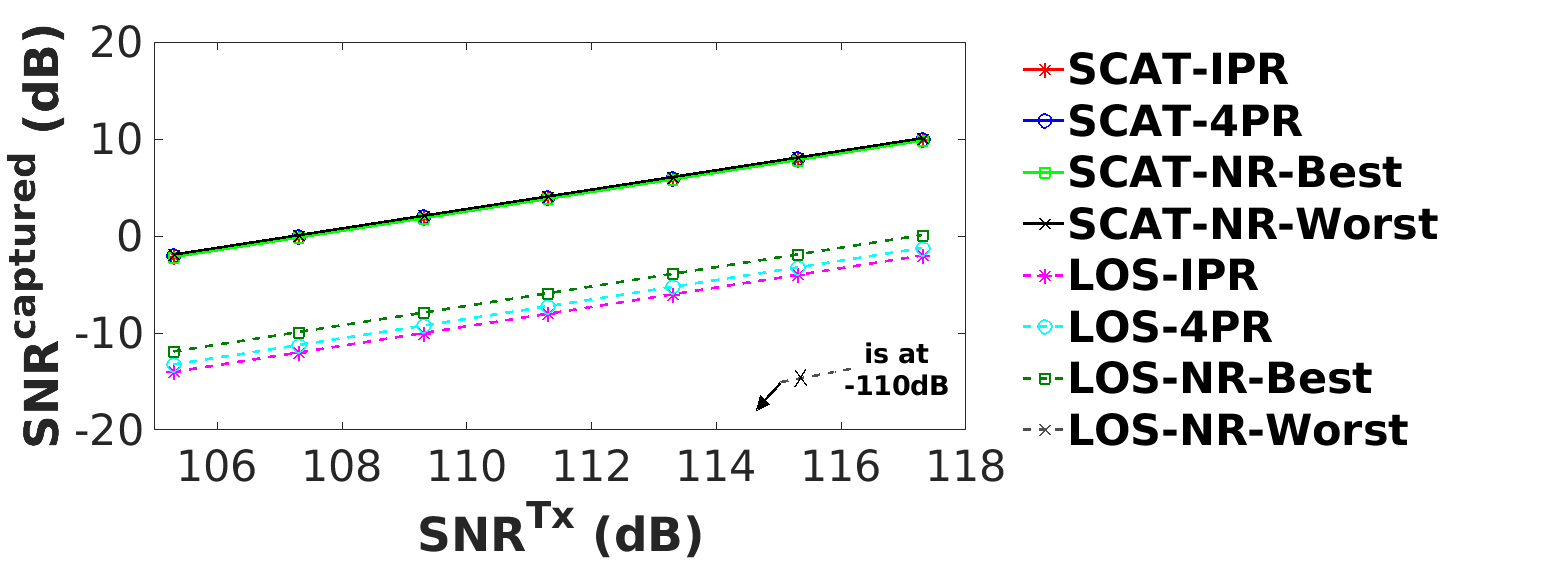}
   \caption{$SNR^{captured}$ by the reader as function of the $SNR^{Tx}$.}
   \label{simulation_SNRt_SNRr}
\end{figure}

Fig.  \ref{simulation_results} shows that the IPR tag attains the best performance. It provides the upper bound performance for this environment configuration. The NR tag, even fixed in the optimum orientation, has a poor outage probability as it is not robust to scattering. Fig. \ref{simulation_SNRt_SNRr} shows the average captured SNR for each configuration as function of $SNR^{Tx}$. As expected, the $SNR^{captured}$ does not depend on the tag polarization (as it is measured when the tag is transparent) and only depends on scattering. However, in LOS, the amount of received power corresponds to the power backscattered by the tag, which depends on the tag polarization. In LOS-NR-Worst configuration, the NR tag and the reader are orthogonal to the source, thus the reader receives close to zero signal from the source and from the tag. 

We observe that the increasing number of reconfigurable polarizations $Npol$ of the tag improves the system performance. In addition, this improvement is boosted by the presence of scatterers. 

Finally, we observe that, even with a limited number of orientations of polarization ($Npol=4$), the performance of the 4PR tag is close to the IPR tag. 

\subsection{Experimental setup}

\begin{figure}
    \centering
    \includegraphics[trim=0cm 0cm 0cm 0.2cm,clip,width=0.9\columnwidth]{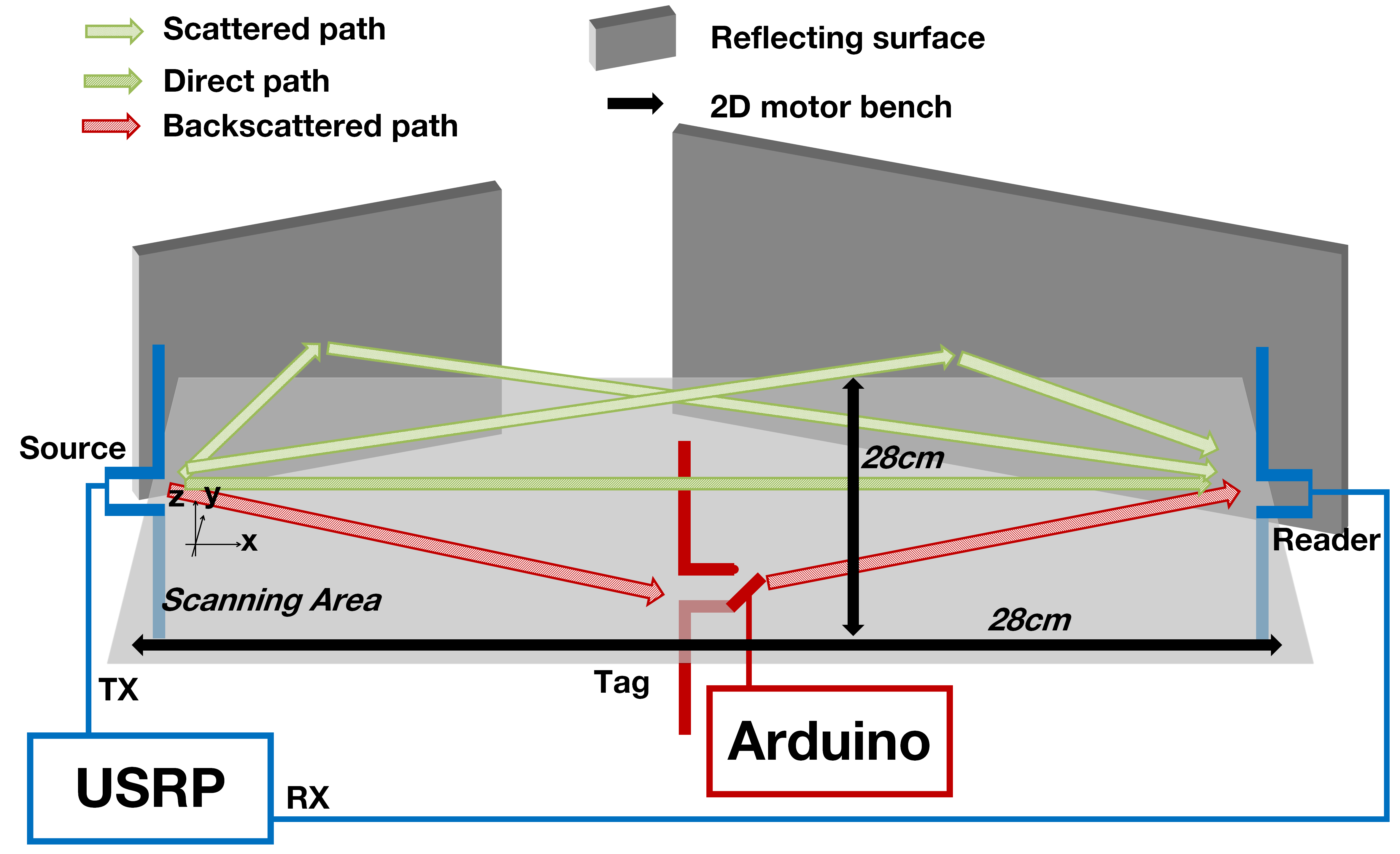}
    \caption{Experimental setup.}
    \label{fig:exp_setup}
\end{figure}{}

\begin{figure}
    \centering
    \includegraphics[trim=0cm 0cm 0cm 0.7cm,clip,width=0.9\columnwidth]{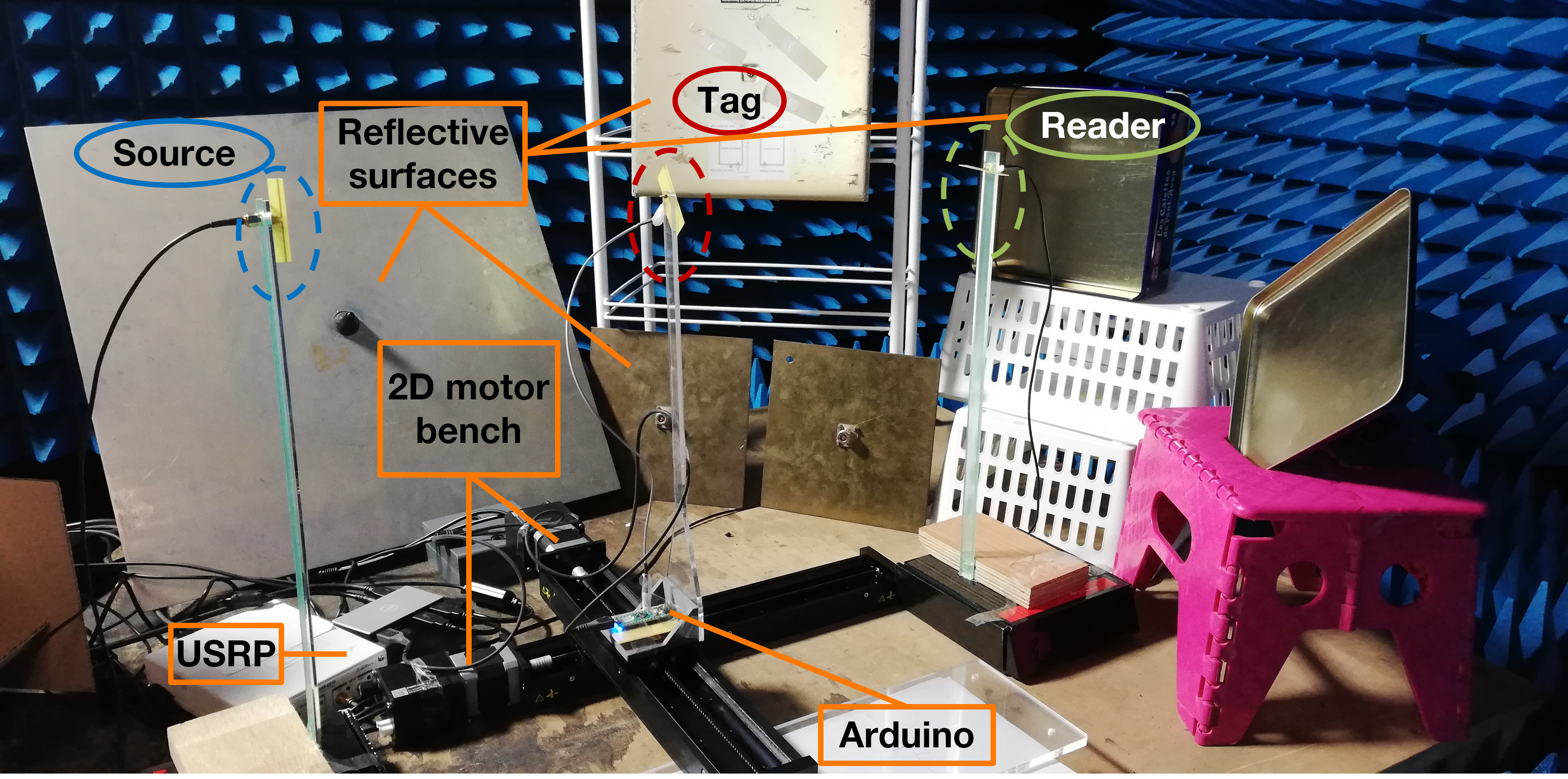}
    \caption{Photo of the experimental setup.}
    \label{fig:photo_exp_setup}
\end{figure}{}

We have shown in previous sections, by simulation, that a PR tag outperforms a NR tag. In this section, we validate this observation experimentally, in a semi-anechoic chamber. In our experiment set-up, we reproduce as closely as possible the conditions modelled in the simulation. Instead of scatterers made of conductive lines, we deploy reflective planes in the environment ($N^{SC}=0$, $N^{RP}=6$). Each of them has different surfaces and is placed randomly around the system, with different orientations and locations, as illustrated in Fig. \ref{fig:exp_setup} and \ref{fig:photo_exp_setup}. Finally, the source is at a shorter distance from the tag and the reader $(0.35m)$ than in the simulation $(100m)$, due to the limited size of the semi-anechoic chamber.

To measure the SNR contrast maps experimentally, the source and the reader are installed at fixed locations. The mechanically rotating dipole antenna of the tag is mounted on two motorized rails with a length of $0.3m$.

Each element is composed of a dipole antenna made of two strands, for a total length of 6.25cm. We connect the source and the reader to two different ports of an USRP B210. We use GNURadio to control the USRP and process the source and received signals. The tag’s antenna is connected to an arduino that controls the impedance connected to the two strands using a diode PIN.

We place the source at the origin of the axis, $(x^S, y^S, z^S)=(0, 0, 0)$, with a vertical orientation, $(\phi^S, \theta^S)=(0, 0)$. Reader is placed such as $(x^R, y^R, z^R)=(0.35, 0, 0)$ and in cross-polarization with the source $(\phi^R, \theta^R)=(90, 90)$. The tag is moved along a linear trajectory perpendicular to the line connecting the source and the reader, such that $x^T=xmin+nx\times step$ and $y^T=ymin+ny\times step$. 
We define $xmin=0.03m$ and $ymin=-0.15m$ so that the tag scans the area between the source and the reader, given the limited range of $0.28m$.
The step of the tag displacement is $step=10^{-2}m$ along x and y axis and we have $nx,ny \in [0,1 \dots, 28]$. 

\subsection{Experimental results}
\begin{figure}[h]
    \centering
  \includegraphics[trim=0cm 0cm 0cm 0cm,clip,width=0.9\columnwidth]{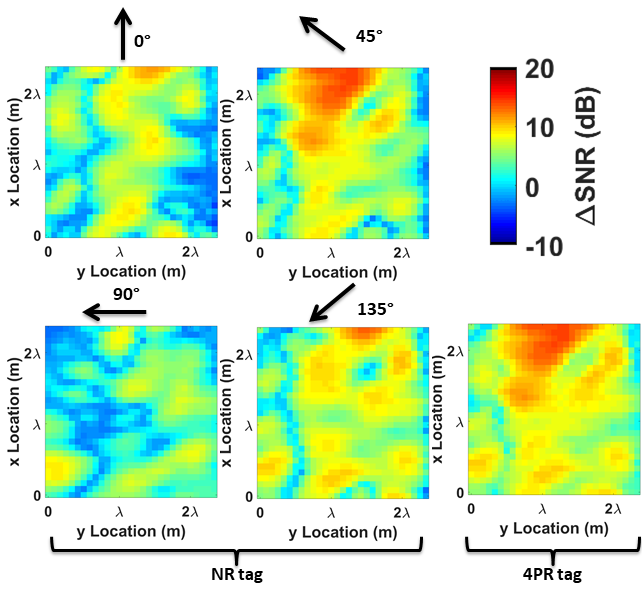}
   \caption{Experimental maps for the 4 fixed orientations of the NR tag ($\phi^T=0^\circ,45^\circ,90^\circ,135^\circ;\theta^T=90^\circ$) and the map for the 4PR tag : the optimum SNR contrast map obtained after selection of the best polarization among the 4 available ones.}
   \label{fig:PC_4states}
\end{figure}

We measure maps of $\Delta P$ and $P^{noise}$ to calculate $\Delta SNR$ maps.
We measure SNR contrast maps for each angle of the 4PR tag (determined in \ref{limited_reconfigurable_tag}). We obtain 4 maps of $\Delta SNR$, that are illustrated in Fig. \ref{fig:PC_4states}. 

We process these maps to determine the map with the optimum SNR contrast. Experimental results show us that selecting among 4 polarization orientations improves general performance of AmB system (Fig. \ref{fig:PC_4states}). Even though the experiment is not a perfect replica of the simulations, experimental and simulation results are consistent: both show that, in a complex environment, $\Delta SNR$ maps depend on the orientation of the polarization of the tag, and that a PR tag is more robust.

\section{Conclusion}
\label{conclusion}

In this paper, we have shown the advantages of using a polarization reconfigurable tag in ambient backscatter systems. Such tag transmits the same message with different linear polarization, to improve the robustness of the communication against direct source-to-reader interference. In this preliminary study, we proposed to use a very simple model of polarization reconfigurable tag to focus our study on polarization. In our model, the compact reconfigurable antenna is modelled by a rotating dipole. We have proposed a simple method to determine the tag orientation that maximizes the performance in a line of sight environment. We have made a first numerical assessment of the improvement in terms of outage probability, obtained with a polarization reconfigurable tag in both a line of sight and richly scattering environments, by simulation and experiments. In particular, we have shown that a realistic 4-polarization reconfigurable tag nearly equals the ideal polarization reconfigurable tag, in performance, with 20 times less polarizations. Further studies will assess a true compact configurable antenna, and account for the joint effects of the reconfigurable antenna gain, directivity and polarization.

\section*{Acknowledgement}
\label{ack}
This work is partially supported by the French Project ANR
Spatial Modulation under grant ANR-15-CE25-0016
(https://spatialmodulation.eurestools.eu/).

\bibliographystyle{IEEEtran}
\bibliography{paper}

\end{document}